\documentclass[aps,nofootinbib,prd,eqsecnum,showpacs,showkeys,preprintnumbers]{revtex4-2}
\usepackage{eurosym}
\usepackage[caption=false]{subfig}
\usepackage{graphicx}
\usepackage{amsmath}
\usepackage{float}
\usepackage{amsfonts}
\usepackage{amssymb}
\usepackage{color}
\usepackage{bm}
\usepackage{mathrsfs}
\usepackage{epstopdf}
\usepackage{url}
\usepackage{footnote}
\usepackage{textcomp}
\usepackage{ulem}
\usepackage{esint}
\usepackage[unicode=true, pdfusetitle,
 bookmarks=true,bookmarksnumbered=false,bookmarksopen=false,
 breaklinks=false,pdfborder={0 0 1},backref=false,colorlinks=false]{hyperref}
\usepackage{multirow}
\usepackage{pifont}

\begin{document}

\title{Constant-roll and primordial black holes in $f(Q,T)$ gravity}
\author{K. El Bourakadi$^{1}$}
\email{k.elbourakadi@yahoo.com}
\author{M. Koussour$^{1}$}
\email{pr.mouhssine@gmail.com}
\author{G. Otalora$^{2}$}
\email{giovanni.otalora@academicos.uta.cl}
\author{M. Bennai$^{1}$}
\email{mdbennai@yahoo.fr}
\author{Taoufik Ouali$^{3}$}
\email{t.ouali@ump.ac.ma}
\date{\today }

\begin{abstract}
In this study, we investigate the consequence of the constant-roll condition
and examine the role of $f(Q,T)$ gravity in the cosmological inflation
process. We analyze the inflationary scenario by calculating modified
Friedmann equations, and giving an alternative technique that enables
relating modified slow-roll parameters to the constant roll parameter $\beta 
$. Considering both chaotic and hilltop models, we calculate the spectral
index and the tensor-to-scalar ratio and compare their compatibility with
Planck's data for different choices of the constant roll parameter $\beta $.
We examine the evolution of primordial black holes in our chosen modified
gravity model taking into account the accretion process and the evaporation
due to Hawking radiation. We compute the evaporation and accretion masses
rate and provide an analytic estimation of the primordial black holes masse
and of the radiation in the $f(Q,T)$ gravity model.
\end{abstract}

\affiliation{$^{1}${\small Quantum Physics and Magnetism Team, LPMC, Faculty of Science
Ben M'sik,}\\
{\small Casablanca Hassan II University, Morocco.}\\} 
\affiliation{$^{2}$Departamento de F\'isica, Facultad de Ciencias,
Universidad de Tarapac\'a, Casilla 7-D, Arica, Chile.\\} 
\affiliation{$^{3}$Laboratory of Physics of Matter and Radiations, \\
 Mohammed first University, BP 717, Oujda, Morocco.\\}
 
\maketitle

\section{Introduction}

\label{secI}

Inflationary cosmology, or simply inflation, is the current widely accepted
paradigm for explaining the physics of very early Universe. In this theory a
quasi-exponential accelerating phase before the radiation decelerating era
is proposed to solve the several long-standing puzzles of the Hot Big-Bang
standard cosmological model \cite%
{Starobinsky:1980te,Guth:1980zm,Albrecht:1982wi,Linde:1981mu}. In addition,
the most fascinating feature of inflation is that it provides a mechanism to
explain the origin of the Cosmic Microwave Background (CMB) temperature
anisotropies and the Large-Scale Structure (LSS) \cite{Weinberg:2008zzc},
and the subsequent phases at which matter appeared for the first time in the
Universe \cite{M0}. The mechanism of inflation is based on the generation of
small quantum fluctuations in the inflaton field, which is the field driving
the accelerated expansion. These quantum fluctuations are amplified in
physical scale during inflation, leading to a Gaussian, scale-invariant and
adiabatic primordial density perturbations \cite%
{Weinberg:2008zzc,Baumann:2018muz}. This information is encoded into the
primordial scalar power spectrum $\mathcal{P}_{s}$ and its scale dependence
is characterized by the scalar spectral index $n_{s}$, which is tightly
constrained by the latest Planck data \cite{Planck:2018jri} to be $%
n_{s}=0.9649\pm 0.0042$ (at $68\%$ C.L.). Furthermore, inflation also
predicts the generation of tensor perturbations as a background of
primordial gravitational waves (PGWs) \cite{D18,M1,M2}. In this case the
amplitude of the primordial tensor power spectrum $\mathcal{P}_{t}$ can be
parameterized in terms of the tensor-to-scalar ratio $r\equiv \mathcal{P}%
_{t}/\mathcal{P}_{s}$ \cite{Maggiore:2018sht}. Although no primordial
gravitational waves have been detected so far, current observations give us
an upper bound on $r$. New data from BICEP/Keck 2021 \cite{BICEP:2021xfz}
have been published, leading to a considerably stronger upper bound on r: $%
r_{0.05}=0.014_{-0.011}^{+0.010}$ ($r_{0.05}<0.036$ (at $95\%$ C.L.)), in
comparison to Planck $2018$ data \cite{Planck:2018jri}.

In general, the dynamics of inflation is based on the slow-roll
approximation. The scalar potential is chosen to be flat such that the
scalar field slowly rolls down the potential \cite{Weinberg:2008zzc}. So,
the equation of the mode functions associated to quantum fluctuations can be
written as a Bessel equation of order $\nu $, and $\nu $ is approximated to
the first order of the Hubble slow-roll parameters $\epsilon _{H}\ll 1$ and $%
\eta _{H}\ll 1$. Thus, the mode functions and the corresponding power
spectra are obtained by applying the so-called Bessel function approximation 
\cite{Wang:1997cw}. However, another alternative that has already been
considered in the literature is the case of ultra-slow roll inflation \cite%
{Tsamis:2003px,Kinney:2005vj,Dimopoulos:2017ged}. The scalar potential is
assumed to be extremely flat, and then the Klein-Gordon equation of motion
of the homogeneous inflaton field is simplified by neglecting the slope
term. Therefore, the friction term is now locked with the acceleration term.
Furthermore, one has that $\eta _{H}\approx 3$ and then the slow-roll
approximation is no longer valid. For instance, the mechanism to generate
primordial black holes within the framework of ultra-slow roll inflation has
been investigated in Refs. \cite{Germani:2017bcs,Di:2017ndc}. Ultra-slow
roll inflation has also been generalized to constant-roll inflation such
that the slow-roll parameter $\eta _{H}$ becomes a constant \cite{B10,B9}
(see also Refs. \cite%
{Motohashi:2017vdc,Motohashi:2017aob,Nojiri:2017qvx,Cicciarella:2017nls,Ito:2017bnn, Gao:2018cpp}%
). Interestingly enough, constant-roll inflation differs from slow-roll
inflation due to its new dynamical features that result in richer physics.
It gives rise to a large local non-Gaussianity and the curvature
perturbation may grow on the super-horizon scales \cite{B10,B9,Namjoo:2012aa}%
.

Primordial Black Holes (PBHs) \cite{I1} started to be considered seriously
after Hawking proposed that PBHs could be formed with Planck-order masses in
the early Universe \cite{I2}. In this direction one would consider that PBHs
formation could be a result of inflation \cite{I3} inhomogeneities from the
early Universe \cite{I4,I5} or the phase transition \cite{I6}. Furthermore,
it was first proposed that PBHs could be a candidate of dark matter (DM) in 
\cite{I7} and reconsidered in \cite{I8,I9} when black holes mergers were
first detected by LIGO \cite{I10}. Hawking found that black holes release
thermal radiation \cite{D3}. As a result, depending on their mass, black
holes evaporate. The faster the PBHs evaporate, the smaller their masses. A
black hole's density, on the other hand, varies inversely with its mass.
Such high densities are required for the formation of lighter black holes,
and such densities are only possible in the early Universe. Primordial black
holes are thus the only black holes with masses small enough to have
vanished by now. A number of observations restrict the mass ranges of PBHs,
for instance, PBHs masses bounded as $\lesssim 10M_{\odot }$ are excluded
from null microlensing searches \cite{I11,I12}, while masses $\gtrsim
10^{2}M_{\odot }$ are excluded from wide binary surveys \cite{I13,I14}. The
essential assumption behind these constraints is that in the early Universe,
PBHs accrete primordial gas and subsequently convert a percentage of the
accreted mass to radiation. The ensuing infusion of energy into the
primordial plasma then influences its thermal and ionization processes,
causing anomalies in the CMB's frequency spectrum and polarization power
spectra \cite{I15,I16}.

In General Relativity (GR), the gravitational interaction is described in
terms of the curvature associated with the Levi-Civita Connection \cite%
{Weinberg:1972kfs}. Moreover, it is well-known that gravity can also be
described in terms of torsion in the context of the so-called Teleparallel
Equivalent of GR, or simply Teleparallel Gravity (TG) \cite%
{Einstein,Unzicker:2005in,Early-papers1,Early-papers2,Early-papers3,Early-papers4,Early-papers5,Early-papers6}%
. TG is a gauge theory for the translation group in which the dynamical
variable is the tetrad field rather than the metric tensor and the torsion
is associated to the Weitzenb\"{o}ck connection that substitutes the
Levi-Civita connection \cite%
{JGPereira2,AndradeGuillenPereira-00,Arcos:2005ec,Pereira:2019woq}.
Additionally, the Lagrangian density of TG is proportional to the torsion
scalar $T$, which is equivalent to the curvature scalar $R$ up to a total
derivative term. Therefore, GR and TG are equivalent at the level of field
equations \cite{Aldrovandi-Pereira-book,Arcos:2005ec}. In all the cases, the
equivalence with GR is only guaranteed for a gravitational action linear in
torsion or decoupled from other fields. On the other hand, as a natural
extension to modified gravity, and similarly to $f(R)$ gravity \cite%
{Clifton:2011jh,Capozziello:2011et,DeFelice:2010aj,Nojiri:2010wj,Nojiri:2006ri}%
, the Lagrangian of TG may be promoted to an arbitrary function $f(T)$ \cite%
{Bengochea:2008gz,Linder:2010py,Li:2011wu} (see also Ref \cite{Cai:2015emx}
and references therein). Also, following the perspective of scalar-tensor
gravity theories \cite{Harrison:1972enf,Clifton:2011jh,Capozziello:2011et},
we can further extend the above theory to $f(T,\phi )$ gravity \cite%
{Hohmann:2018rwf,Gonzalez-Espinoza:2020azh}. These novel modified gravity
theories have shown a rich structure by explaining the dynamics of dark
energy \cite{Skugoreva:2014ena,Gonzalez-Espinoza:2020jss,
Gonzalez-Espinoza:2021mwr,Leon:2022oyy,Duchaniya:2022fmc} and inflation \cite%
{Gonzalez-Espinoza:2020azh,Gonzalez-Espinoza:2021qnv,Leyva:2021fuo}.

Thus, in the context of GR, without considering any modification to gravity,
one can see that there exist at least two equivalent descriptions. The
curvature representation, in which the torsion and nonmetricity vanish, and
the teleparallel representation, in which the curvature and nonmetricity
vanish. But, recently a third equivalent representation has also been
proposed. In this approach, called Symmetry Teleparallel Gravity (STG), the
curvature and torsion vanish, and then the gravitational interaction is
described by the nonmetricity $Q$ of the metric \cite{Nester:1998mp}. STG
has also been further developed to tackle modify gravity in the context of $%
f(Q)$ gravity theory, also known as nonmetric gravity \cite%
{BeltranJimenez:2017tkd}. Recently, interest in this theory has increased
rapidly due to its novel geometrical and physical features \cite%
{Adak:2004uh,Adak:2006rx,Adak:2008gd,BeltranJimenez:2015pnp,Adak:2018vzk,Soudi:2018dhv,Jarv:2018bgs,Harko:2018gxr,Lazkoz:2019sjl,BeltranJimenez:2019tme,BeltranJimenez:2019esp}%
, $f(Q)$ gravity is also used to explain cosmic acceleration \cite%
{K1,K2,K3,K4,K5}. Moreover, another proposed theory of modified gravity
based on nonmetricity is $f(Q,T)$ gravity \cite{A1}, where $T$ is the trace
of the matter energy-momentum tensor. Thus, it is an extension of $f(Q)$
gravity where a direct coupling between nonmetricity and matter has been
assumed. In fact, one can verify that the divergence of the matter
energy-momentum tensor does not vanish by taking the covariant derivative of
the modified field equations. Several different aspects of $f(Q,T)$ gravity
have been studied, among these we have observational constraints \cite{A2},
energy conditions \cite{Arora:2020iva,Arora:2020met}, cosmological solutions 
\cite{Godani:2021mld}, phase-space dynamics \cite{Pati:2021ach}, and
cosmological inflation \cite{Shiravand:2022ccb}.

The plan of the paper is the following: In Section \ref{secII}, we present
an introduction to $f\left( Q,T\right) $ gravity. In Section \ref{secIII},
we study the dynamics of constant-roll inflation within the context of $%
f(Q,T)$ gravity. In Section \ref{secIV}, we investigate the generation of
primordial black holes in $f(Q,T)$ gravity. Finally, in Section \ref{secV},
we summarize the results obtained.


\section{Basic formalism in $f\left( Q,T\right) $ gravity}

\label{secII}

In the so-called Weyl-Cartan geometry, gravitational effects are caused not
only by a variation of the direction of a vector in parallel transport, but
also by a variation in its length. The variation in the length of a vector
is described geometrically by what is called non-metricity, and it is
mathematically described as the covariant derivative of the metric tensor
which is a generalization of the gravitational potential. In this case,
general connexion $\Sigma _{\ \mu \nu }^{\alpha }$ is written in terms of
all contributions as curvature, torsion and non-metricity of spacetime \cite%
{Tom},%
\begin{equation}
\Sigma _{\ \mu \nu }^{\alpha }=\Gamma _{\ \mu \nu }^{\alpha }+C_{\ \mu \nu
}^{\alpha }+L_{\ \mu \nu }^{\alpha },  \label{2a}
\end{equation}%
where $\Gamma _{\ \mu \nu }^{\alpha }$ is the famous Levi-Civita connection
of the metric tensor $g_{\mu \nu }$ in GR given by,%
\begin{equation}
\Gamma _{\ \mu \nu }^{\alpha }\equiv \frac{1}{2}g^{\alpha \lambda }(g_{\mu
\lambda ,\nu }+g_{\lambda \nu ,\mu }-g_{\mu \nu ,\lambda }),  \label{2b}
\end{equation}%
here $C_{\ \mu \nu }^{\alpha }$ represents the Contortion tensor expressed
as,%
\begin{equation}
C_{\ \mu \nu }^{\alpha }\equiv \frac{1}{2}(\tilde{T}_{\ \mu \nu }^{\alpha }+%
\tilde{T}_{\mu \ \nu }^{\ \alpha }+\tilde{T}_{\nu \ \mu }^{\ \alpha })=-C_{\
\nu \mu }^{\alpha },  \label{2c}
\end{equation}%
and $L_{\ \mu \nu }^{\alpha }$ is the Disformation tensor given\ by,%
\begin{equation}
L_{\ \mu \nu }^{\alpha }\equiv \frac{1}{2}(Q_{\ \mu \nu }^{\alpha }-Q_{\mu \
\nu }^{\ \alpha }-Q_{\nu \ \mu }^{\ \alpha })=L_{\ \nu \mu }^{\alpha },
\label{2d}
\end{equation}%
where $\tilde{T}_{\ \mu \nu }^{\alpha }$ and $Q_{\alpha \mu \nu }$ in Eqs. %
\eqref{2c} and \eqref{2d} are the torsion tensor and the non-metricity
tensor, respectively. In non-flat space-time, the geodesic structure is
designated by the connection form. In addition, in Einstein's GR, the
presumption of a non-torsion and metric compatible connection leads to the
so-called Levi-Civita connection, connected to the metric tensor and its
first derivatives. It is possible to define two tensors of order 3 related
to the antisymmetric part of $\Sigma _{\ \mu \nu }^{\alpha }$ and the
covariant derivative of the metric tensor as,%
\begin{equation}
\tilde{T}_{\ \mu \nu }^{\alpha }\equiv 2\Sigma _{\ \left[ \mu \nu \right]
}^{\alpha },  \label{2e}
\end{equation}%
and%
\begin{equation}
Q_{\alpha \mu \nu }\equiv \nabla _{\alpha }g_{\mu \nu }=\partial _{\alpha
}g_{\mu \nu }-g_{\nu \sigma }\Sigma {^{\sigma }}_{\mu \alpha }-g_{\sigma \mu
}\Sigma {^{\sigma }}_{\nu \alpha }\neq 0.  \label{2f}
\end{equation}

Depending on the proposed connection, different theories of gravity that can
be extracted. In this work, we will consider the modified Einstein-Hilbert
action in $f(Q,T)$ extended symmetric teleparallel gravity expressed as \cite%
{A0,A1},%
\begin{equation}
S=\int \sqrt{-g}d^{4}x\left( \frac{1}{2\kappa }f(Q,T)+L_{m}\right) ,
\label{eqn1}
\end{equation}%
where $f(Q,T)$ is a general function of the non-metricity scalar $Q$ and the
trace of the energy-momentum tensor $T$,\ $\kappa =1/M_{p}^{2},$ $g$ is the
determinant of the metric tensor $g_{\mu \nu }$ i.e. $g=\det \left( g_{\mu
\nu }\right) $,\ and $L_{m}$ is the Lagrangian function for the matter
fields.\ The non-metricity scalar $Q$ is defined as,%
\begin{equation}
Q\equiv -g^{\mu \nu }(L_{\,\,\,\alpha \mu }^{\beta }L_{\,\,\,\nu \beta
}^{\alpha }-L_{\,\,\,\alpha \beta }^{\beta }L_{\,\,\,\mu \nu }^{\alpha }).
\label{eqn2}
\end{equation}

The trace of the non-metricity tensor is obtained as, 
\begin{equation}
Q_{\beta }=g^{\mu \nu }Q_{\beta \mu \nu }\qquad \widetilde{Q}_{\beta
}=g^{\mu \nu }Q_{\mu \beta \nu }.  \label{eqn5}
\end{equation}

Further, the superpotential tensor $P_{\,\,\,\mu \nu }^{\beta }$ (or
non-metricity conjugate) is related to the non-metricity scalar $Q$ as, 
\begin{equation}
P_{\,\,\,\mu \nu }^{\beta }=-\frac{1}{2}L_{\,\,\,\mu \nu }^{\beta }+\frac{1}{%
4}(Q^{\beta }-\widetilde{Q}^{\beta })g_{\mu \nu }-\frac{1}{4}\delta _{(\mu
}^{\beta }Q_{\nu )}.  \label{eqn6}
\end{equation}

By using the definition above, the non-metricity scalar in terms of the
superpotential tensor is given by, 
\begin{equation}
Q=-Q_{\beta \mu \nu }P^{\beta \mu \nu }.  \label{eqn7}
\end{equation}

As usual, we define of the energy-momentum tensor of the matter fields by 
\begin{equation}
T_{\mu \nu }=-\frac{2}{\sqrt{-g}}\dfrac{\delta (\sqrt{-g}L_{m})}{\delta
g^{\mu \nu }},  \label{eqn8}
\end{equation}%
and 
\begin{equation}
\Theta _{\mu \nu }=g^{\alpha \beta }\frac{\delta T_{\alpha \beta }}{\delta
g^{\mu \nu }}.  \label{eqn9}
\end{equation}

So that the variation of energy-momentum tensor with respect to the metric
tensor $g_{\mu \nu }$\ read as, 
\begin{equation}
\frac{\delta \,\left( g^{\,\mu \nu }\,T_{\,\mu \nu }\right) }{\delta
\,g^{\,\alpha \,\beta }}=T_{\,\alpha \beta }+\Theta _{\,\alpha \,\beta }.
\label{eqn10}
\end{equation}

Now, varying the gravitational action (\ref{eqn1}) with respect to metric
tensor $g_{\mu \nu }$, the fleld equations of $f(Q,T)$ gravity can be
derived as, 
\begin{equation}
-\frac{2}{\sqrt{-g}}\nabla _{\beta }\left( f_{Q}\sqrt{-g}P_{\,\,\,\,\mu \nu
}^{\beta }\right) -\frac{1}{2}fg_{\mu \nu }+f_{T}(T_{\mu \nu }+\Theta _{\mu
\nu })-f_{Q}(P_{\mu \beta \alpha }Q_{\nu }^{\,\,\,\beta \alpha
}-2Q_{\,\,\,\mu }^{\beta \alpha }P_{\beta \alpha \nu })=\kappa T_{\mu \nu }.
\label{eqn11}
\end{equation}

Here, $f_{Q}=\frac{df\left( Q,T\right) }{dQ}$, $f_{T}=\frac{df\left(
Q,T\right) }{dT}$, and $\nabla _{\beta }$\ denotes the covariant derivative.
Now, we consider a homogeneous and spatially flat
Friedmann-Lemaitre-Robertson-Walker (FLRW) metric described by the line
element 
\begin{equation}
ds^{2}=-N^{2}\left( t\right) dt^{2}+a^{2}(t)\left(
dx^{2}+dy^{2}+dz^{2}\right) ,  \label{eqn12}
\end{equation}%
where $a\left( t\right) $ is the scale factor of the Universe, depending
only on the cosmic time $t$ (which is scaled to be unity at the present
time, i.e. $a_{0}=1$) and $N\left( t\right) $\ is the lapse function
regarded to be $1$ in the standard model. The rates of expansion and
dilation are fixed as $H\equiv \frac{\overset{.}{a}}{a}$ and $D\equiv \frac{%
\overset{.}{N}}{N}$, respectively. Hence, the corresponding non-metricity
scalar is given by $Q=6\frac{H^{2}}{N^{2}}$. In our current analysis, we
suppose that the content of the Universe as a perfect non-viscosity fluid
for which the energy-momentum tensor is given by 
\begin{equation}
T_{\nu }^{\mu }=diag\left( -\rho ,p,p,p\right) ,  \label{eqn13}
\end{equation}%
where $p$ is the perfect non-viscosity fluid pressure and $\rho $ is the
energy density of the Universe. Hence, for the tensor $\Theta _{\nu }^{\mu }$
the expression is obtained as $\Theta _{\nu }^{\mu }=diag\left( 2\rho
+p,-p,-p,-p\right) $. Taking into account the case as $N=1$, the Einstein
field equations using the line element (\ref{eqn12}) are given by, 
\begin{equation}
\kappa \rho =\frac{f}{2}-6FH^{2}-\frac{2E}{1+E}\left( \overset{.}{F}H+F%
\overset{.}{H}\right)   \label{eqn14}
\end{equation}%
and%
\begin{equation}
\kappa p=-\frac{f}{2}+6FH^{2}+2\left( \overset{.}{F}H+F\overset{.}{H}\right)
.  \label{eqn15}
\end{equation}%
where we used $Q=6H^{2}$ and $\left( \text{\textperiodcentered }\right) $
represents a derivative with respect to cosmic time $\left( t\right) $. In
this case, $F\equiv f_{Q}$ and $\kappa E\equiv f_{T}$ represent
differentiation with respect to $Q$ and $T$ respectively. The evolution
equation for the Hubble parameter $H$ can be derived by combining Eqs. (\ref%
{eqn14}) and (\ref{eqn15}) as, 
\begin{equation}
\overset{.}{H}+\frac{\overset{.}{F}}{F}H=\frac{\kappa }{2F}\left( 1+E\right)
\left( \rho +p\right) .  \label{eqn16}
\end{equation}

Einstein's field equations (\ref{eqn14}) and (\ref{eqn15}) can be regarded
as extended symmetric teleparallel equivalents to standard Friedmann's
equations with supplementary terms from the non-metricity of space-time and
the trace of the energy-momentum tensor $T$ which behaves as an effective
component. Therefore, the effective energy density $\rho _{eff}$\ and
effective pressure $p_{eff}$\ are determined by, 
\begin{equation}
3H^{2}=\kappa \rho _{eff}=\frac{f}{4F}-\frac{\kappa }{2F}\left[ \left(
1+E\right) \rho +Ep\right] ,  \label{eqn17}
\end{equation}%
\begin{equation}
2\dot{H}+3H^{2}=-\kappa p_{eff}=\frac{f}{4F}-\frac{2\dot{F}H}{F}+\frac{%
\kappa }{2F}\left[ \left( 1+E\right) \rho +\left( 2+E\right) p\right] .
\label{eqn18}
\end{equation}%
Taking into account Eqs. (\ref{eqn16}) and (\ref{eqn17}) one gets

\begin{equation}
\rho =\frac{f-12H^{2}F}{2\kappa \left[ \left( 1+\omega \right) E+1\right] }.
\label{gh}
\end{equation}

\section{Inflationary scenario in $f(Q,T)$ gravity}

\label{secIII}

\subsection{Slow-roll and constant-roll in the standard cosmology}

The simple inflation model is described with an isotropic and homogeneous
scalar field known as the inflaton. The dynamics of such inflaton field can
be defined via the Lagrangian%
\begin{equation}
L_{m}=-\frac{1}{2}g^{\mu \nu }\partial _{\mu }\phi \partial _{\nu }\phi
-V(\phi ).  \label{L}
\end{equation}%
From the Lagrangian of the scalar field $\phi $\ the energy-momentum tensor
is defined as\ 
\begin{equation}
T_{\mu \nu }\equiv -\frac{2}{\sqrt{-g}}\frac{\delta \left( \sqrt{-g}%
L_{m}\right) }{\delta g^{\mu \nu }}=g_{\mu \nu }L_{m}-2\frac{\partial L_{m}}{%
\partial g^{\mu \nu }}.  \label{T}
\end{equation}%
Considering the inflaton field as a perfect fluid with an equation of state $%
\omega =p_{\phi }/\rho _{\phi },$\ knowing that $\rho _{\phi }$ and $p_{\phi
}$\ are the energy density and the pressure of the inflaton field, the
energy-momentum tensor in Eq. \ref{T} gives 
\begin{equation}
\rho _{\phi }=\frac{1}{2}\dot{\phi}^{2}+V(\phi ),  \label{eqp}
\end{equation}%
and%
\begin{equation}
p_{\phi }=\frac{1}{2}\dot{\phi}^{2}-V(\phi ).  \label{eqgho}
\end{equation}%
Here the dot represents the derivative with respect to the cosmic time. The
equation of state parameter is then written as%
\begin{equation}
\omega =\frac{p_{\phi }}{\rho _{\phi }}=\frac{\frac{1}{2}\dot{\phi}%
^{2}-V(\phi )}{\frac{1}{2}\dot{\phi}^{2}+V(\phi )}.
\end{equation}%
Taking into account the Einstein field equations in addition to Eq. (\ref%
{eqp}) and\ (\ref{eqgho}), the Friedmann equations are easily obtained as 
\begin{eqnarray}
H^{2} &=&\frac{\kappa }{3}\left[ \frac{\dot{\phi}^{2}}{2}+V(\phi )\right] ,
\label{H2} \\
H^{2}+\dot{H} &=&-\frac{\kappa }{3}\left[ \dot{\phi}^{2}-V(\phi )\right] , \\
\dot{H} &=&-\frac{\kappa }{2}\dot{\phi}^{2},  \label{H}
\end{eqnarray}%
here $H\equiv \dot{a}/a$\ is the Hubble parameter. The so called
Klein-Gordon (KG) equation, by taking the time derivative of Eq. (\ref{H2})
and by taking into considering Eq. (\ref{H}), can be obtained 
\begin{equation}
\ddot{\phi}+3H\dot{\phi}+V^{\prime }=0,  \label{KG}
\end{equation}%
here the prime denotes the derivative with respect to the $\phi $-field. The
accelerated expansion of the inflationary phase must last for enough time
for a successful period where the Hubble radius decreased over time. For
this, the slow-roll parameters are defined as \cite{B1} 
\begin{eqnarray}
\epsilon  &=&-\frac{\dot{H}}{H^{2}},  \label{eps1} \\
\eta  &=&\frac{\dot{\epsilon}}{H\epsilon }\approx -\frac{\ddot{H}}{2\dot{H}H}%
.  \label{eps2}
\end{eqnarray}%
An additional parameter used to study the period of inflation is the
e-folding number which describes the rate of the Univers expansion during
this phase \cite{B3,B4} 
\begin{equation}
N\equiv \ln \left( \frac{a_{end}}{a}\right) =\int_{t}^{t_{end}}Hdt,
\end{equation}%
the index $"_{end}"$ denotes the time when inflation ended. Inflation will
continue as long as $\epsilon <1$\ to solve the standard cosmological
problems. However, at the end of inflation, the slow-roll parameter must
reach $\epsilon =1.$ The slow roll parameters and the curvature
perturbations are related by the spectral index $n_{s}$, and the ratio of
tensor to scalar perturbations $r$, in the following way \cite{B5}%
\begin{eqnarray}
n_{s}-1 &=&-6\epsilon +2\eta ,  \label{ns} \\
r &=&16\epsilon .  \label{r}
\end{eqnarray}%
\newline
The Planck data provides strong constraints on $n_{s}$ and $r$ parameters.
In fact, any inflationary model predict such parameters can be tested to
decide whether it can be ruled out or not \cite{B6,B7}. In addition to the
slow roll conditions, one may consider a type of constant roll of the scalar
field, which is expressed as follows%
\begin{equation}
\frac{\ddot{\phi}}{\dot{\phi}}=\beta H.
\end{equation}%
In Ref. \cite{B8} they considered two slow-roll stages, separated by a
constant-roll stage that imprints an alternative dynamic to the scalar field 
\cite{B9}, where the constant $\beta $ determines the deviation from a flat
potential. When $\beta \simeq 0$, the slow roll inflation is recovered,
whereas $\beta =0$ corresponds to the \textquotedblleft ultra slow
roll\textquotedblright\ inflation \cite{B10,B11}.

In order to construct the idea of constant roll inflation this paper aims to
use the condition $\epsilon \ll 1$ that leads to 
\begin{equation}
\dot{\phi}^{2}\ll V(\phi ).
\end{equation}%
which simplifies Eq. (\ref{eps2}) as%
\begin{equation*}
\eta \approx -\frac{\left\vert \ddot{\phi}\right\vert }{H\left\vert \dot{\phi%
}\right\vert }.
\end{equation*}%
However, for a successful constant roll scenario, we consider the expression 
$\ddot{\phi}=\dot{\phi}\beta H$ that simplifies the KG equation as

\begin{equation}
\dot{\phi}H\left( 3+\beta \right) +V^{\prime }=0.
\end{equation}
\ In the next section, we will investigate the effect of the constant roll
stage on the inflationary parameters by deriving $H,$\ $\dot{H},$ $\dot{\phi}
$\ and $\ddot{\phi}$\ from an$\ f(Q,T)$ gravity perspective.

\subsection{ Inflation in $f(Q,T)$ Gravity}

In the context of cosmic inflation, we consider the model $f(Q,T)=\alpha
Q+\sigma T$ \cite{A1}, with $\alpha =F\neq 0$, $\sigma $\ is an arbitrary
constant, and $T=-\rho +3p$.\ It is crucial to keep in mind that $\sigma =0$%
\ and $\alpha =-1$\ is equivalent to the case of the GR theory. Furthermore,
the theory is reduced to $f(Q,T)=-Q$\ for $\alpha =-1$\ at $\sigma \neq 0$\
and $T=0$\ which is equivalent to GR as well. By taking Eqs.\ (\ref{eqn16}),
(\ref{eqn17}), (\ref{eqn18}) and (\ref{gh}),\ we obtain the following
equations%
\begin{eqnarray}
\dot{H} &=&\frac{\left( \kappa +\sigma \right) \rho \left( 1+\omega \right) 
}{2\alpha },  \label{Hp} \\
3H^{2} &=&\kappa \rho _{eff}=\frac{\left[ \left( \omega -3\right) \sigma
-2\kappa \right] \rho }{2\alpha },  \label{EqiX} \\
2\dot{H}+3H^{2} &=&-\kappa p_{eff}=\frac{\rho \left[ \left( 3\omega
-1\right) \sigma +2\omega \kappa \right] }{2\alpha },  \label{EqiY} \\
\rho &=&\frac{-6\alpha H^{2}}{\sigma \left( 3-\omega \right) +2\kappa }.
\end{eqnarray}%
Aside from the equation of state that corresponds to the inflationary phase,
we can derive an effective equation of state for this model as%
\begin{equation}
\omega _{eff}=\frac{p_{eff}}{\rho _{eff}}=-1-\frac{2\left( \kappa +\sigma
\right) \left( 1+\omega \right) }{\left( \omega -3\right) \sigma -2\kappa }.
\end{equation}%
It is clear that from the effective equation of state above, we can have an
accelerated de Sitter expansion as long as $\left( \left( \omega -3\right)
\sigma -2\kappa \neq 0\right) $\ and $\left( \omega =-1\right) .$\
Considering the trace of the energy-momentum tensor as $T=\dot{\phi}^{2}-4V$%
, the cosmological inflation in the context of $f(Q,T)$ gravity, from Eqs. (%
\ref{L}), Eqs.(\ref{eqn17}) and (\ref{eqn18}) one obtains%
\begin{eqnarray}
\rho _{eff} &=&-\frac{\left( \kappa +\sigma \right) \dot{\phi}^{2}+2V(\phi
)\left( \kappa +2\sigma \right) }{2\kappa \alpha }, \\
p_{eff} &=&-\frac{\left( \kappa +\sigma \right) \dot{\phi}^{2}-2V(\phi
)\left( \kappa +2\sigma \right) }{2\kappa \alpha }.
\end{eqnarray}%
On the other hand, the modified Friedmann equations can be given as%
\begin{eqnarray}
\frac{3}{2}H^{2} &=&-\frac{\left( \kappa +\sigma \right) \dot{\phi}%
^{2}+2V(\phi )\left( \kappa +2\sigma \right) }{4\alpha },  \label{pp} \\
2\dot{H}+\frac{3}{2}H^{2} &=&\frac{3\left( \kappa +\sigma \right) \dot{\phi}%
^{2}-2V(\phi )\left( \kappa +2\sigma \right) }{4\alpha }.  \label{p}
\end{eqnarray}%
Taking into account the previous two equations one gets%
\begin{equation}
\dot{H}=\frac{\dot{\phi}^{2}}{2\alpha }\left( \kappa +\sigma \right) ,
\label{H.}
\end{equation}%
the effective equation of state parameter can be obtained as%
\begin{equation}
\omega _{eff}=-\frac{\left( \kappa +\sigma \right) \dot{\phi}^{2}-2V(\phi
)\left( \kappa +2\sigma \right) }{\left( \kappa +\sigma \right) \dot{\phi}%
^{2}+2V(\phi )\left( \kappa +2\sigma \right) }.
\end{equation}%
The derivation of Eq.(\ref{pp}) and using Eq. (\ref{H.}), we obtain the
modified equation%
\begin{equation}
\ddot{\phi}\left( \kappa +\sigma \right) +3H\dot{\phi}\left( \kappa +\sigma
\right) +V^{\prime }\left( \kappa +2\sigma \right) =0.
\end{equation}%
From the constant roll condition $\ddot{\phi}=\beta H\dot{\phi},$\ we obtain%
\begin{equation}
\dot{\phi}=-\frac{V^{\prime }\left( \kappa +2\sigma \right) }{H\left( \beta
+3\right) \left( \kappa +\sigma \right) }.  \label{phip}
\end{equation}%
However, one may consider the case where the bound $\left( \kappa +\sigma
\right) \dot{\phi}^{2}\ll V(\phi )\left( \kappa +2\sigma \right) $ takes
place and from Eq.(\ref{pp}), Eq.(\ref{eps1}) and Eq.(\ref{H.}) to obtain
the following slow roll parameter%
\begin{equation}
\epsilon \approx \frac{3\left( \kappa +\sigma \right) \dot{\phi}^{2}}{%
2\left( \kappa +2\sigma \right) V}=\frac{-9\alpha }{2\left( \beta +3\right)
^{2}\left( \kappa +\sigma \right) }\left( \frac{V^{\prime }}{V}\right) ^{2},
\label{e}
\end{equation}%
deriving Eq.(\ref{H.}) and using Eq.(\ref{eps2}), the second slow roll
parameter can simply be given as%
\begin{equation}
\eta \approx -\frac{\ddot{H}}{2\dot{H}H}=-\frac{\left\vert \ddot{\phi}%
\right\vert }{H\left\vert \dot{\phi}\right\vert }=\frac{-3\alpha }{\left(
\beta +3\right) \left( \kappa +\sigma \right) }\left( \frac{V^{\prime \prime
}}{V}\right) .  \label{n}
\end{equation}%
Eqs. (\ref{e}) and (\ref{n}) are obtained by assuming that the constant roll
stage succeeds the slow roll one.

\subsection{Chaotic potential}

Let us consider the simplest model known as chaotic potential, given by the
form%
\begin{equation}
V\left( \phi \right) =\frac{1}{2}m^{2}\phi ^{2},
\end{equation}%
the parameters $\epsilon $ and $\eta $\ are expressed as a function of the
constant roll parameter\ $\beta $\ in the following way%
\begin{eqnarray}
\epsilon &=&\frac{3}{2}\frac{1-n_{s}}{6-\beta }, \\
\eta &=&\left( \frac{3+\beta }{6-\beta }\right) \frac{1-n_{s}}{2},
\end{eqnarray}%
here we should note that $\left( 6-\beta \right) \neq 0.$\ To obtain Eqs. (%
\ref{e}) and (\ref{n}), we should develop a little bit the transition from
slow roll stage to a constant roll one, for the chaotic potential we see
that the behavior of the inflationary phase is determined through the
constant roll parameter $\beta .$ 
\begin{figure}[H]
\centering
\includegraphics[width=16cm]{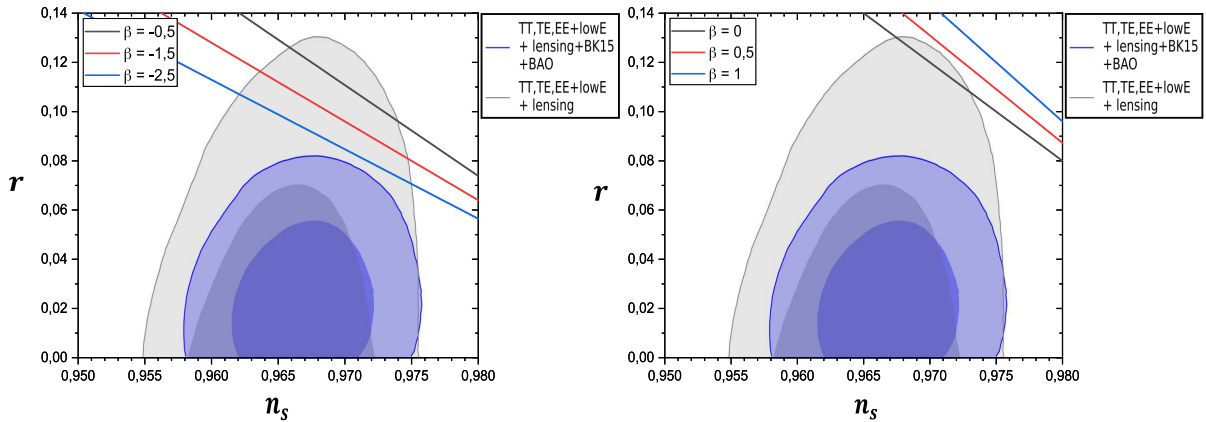}
\caption{$r$ as a function of $n_{s}$ for a chaotic potential in modified
symmetric teleparallel gravity, light and dark blue regions are constraints
in combination with CMB lensing reconstruction and BAO from Planck data. on
the right panel we choose three different values of $\protect\beta $, the
black line for $\protect\beta =0$, the red line represents $\protect\beta %
=0.5,$ and the blue line for $\protect\beta =1$. While on the left one, We
choose three different values of $\protect\beta $, the black line for $%
\protect\beta =-0.5$, the red line represents $\protect\beta =-1.5,$ and the
blue line for $\protect\beta =-2.5$. }
\label{fig:1}
\end{figure}

Fig. \ref{fig:1} show the decreasing behavior of the tensor-to-scalar ratio, 
$r$, with respect to the spectral index, $n_{s}$ for the chaotic potential
in our chosen $f(Q,T)$ gravity model. The results show a good consistency
for a specific range of the constant roll parameter $\beta $ with the latest
observations from Planck data. Furthermore, the case with $\beta >0$
provides inconsistent results with Planck's data. However, the constant roll
parameter must be bounded as $\beta \leq 0$ in order to produce consistent
observational parameters with recent results.

\subsection{Hilltop inflation}

Hilltop inflation is classified among small-field models. This potential
naturally involves eternal inflation, that raises the question of initial
conditions, which is a problem in most inflation models \cite{C1}. Hilltop
potential is given by \cite{C2}%
\begin{equation}
V\left( \phi \right) =M^{4}\left[ 1-\left( \frac{\phi }{\mu }\right) ^{p}%
\right] ,
\end{equation}%
this model of inflation has two free parameters $M$ and $%
{\mu}%
$ plus the parameter $p$\ that will be given specific values at the end of
this section. The e-folds number during inflation using Eqs. (\ref{phip})
and (\ref{pp}) is given by%
\begin{equation}
N\equiv \int_{\phi _{k}}^{\phi _{end}}\frac{H}{\dot{\phi}}d\phi \approx -%
\frac{(\beta +3)\left( \kappa +\sigma \right) }{3\alpha }\int_{\phi
_{end}}^{\phi _{k}}\frac{V}{V^{\prime }}d\phi ,
\end{equation}%
\newline
considering $\Phi =\phi /\mu $\ and $p>2,$\ the expression of inflationary
e-folds number is given by%
\begin{equation*}
N=-\frac{\mu ^{2}}{2p}\frac{(\beta +3)\left( \kappa +\sigma \right) }{%
3\alpha }\left[ \Phi _{k}^{2}-\Phi _{end}^{2}+\frac{2}{p-2}\Phi _{k}^{2-p}-%
\frac{2}{p-2}\Phi _{end}^{2-p}\right] ,
\end{equation*}%
here the subscripts $"_{k}"$\ and $"_{end}"$ mean the time the pivot scale
crossed outside the horizon and the end of inflation respectively. Since the
slow roll parameters are defined by the value of the field at the horizon
crossing $\Phi _{k},$ for this model $\epsilon $\ and $\eta $\ are
calculated as follows 
\begin{eqnarray}
\epsilon &=&-\frac{9\alpha }{2(\beta +3)^{2}\left( \kappa +\sigma \right) }%
\frac{p^{2}}{\mu ^{2}}\frac{\Phi _{k}^{2\left( p-1\right) }}{\left( 1-\Phi
_{k}^{p}\right) ^{2}}, \\
\eta &=&\frac{3\alpha }{(\beta +3)\left( \kappa +\sigma \right) }\frac{%
p\left( p-1\right) }{\mu ^{2}}\frac{\Phi _{k}^{p-2}}{1-\Phi _{k}^{p}}.
\end{eqnarray}%
\ Furthermore, we can compute the tensor-to-scalar ratio as a function of
the inflationary e-folds $N$, the constant-roll parameter $\beta $ for $p>2$
in the following way,%
\begin{equation}
r=\frac{4p}{(\beta +3)}\frac{1}{N}.  \label{r(N)}
\end{equation}%
Taking into consideration Eq. (\ref{r}), we can study the behavior of $r$ as
a function of the spectral index $n_{s}$, $\beta $\ and $p$\ parameters
using the first slow roll parameter $\epsilon $\ given as 
\begin{eqnarray}
\epsilon &=&\frac{3p\left( 1-n_{s}\right) }{18p-4\left( p-1\right) \left(
\beta +3\right) }, \\
\eta &=&\frac{2\left( p-1\right) \left( \beta +3\right) \left(
1-n_{s}\right) }{-18p+4\left( p-1\right) \left( \beta +3\right) }.
\end{eqnarray}%
\ 
\begin{figure}[H]
\centering
\includegraphics[width=12cm]{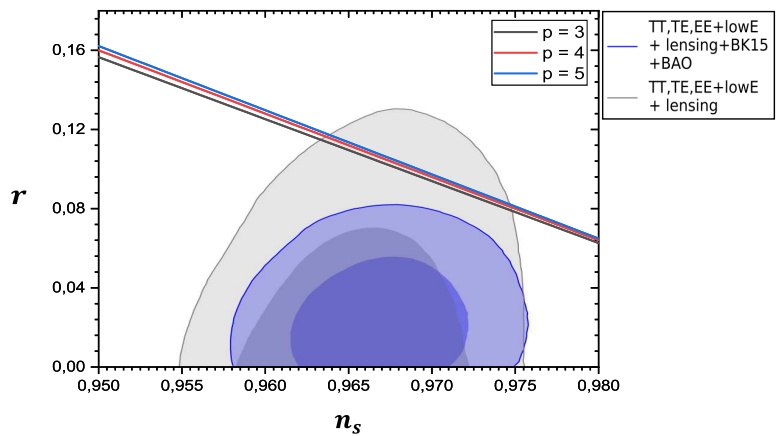}
\caption{$r$ as a function of $n_{s}$ for the hilltop potential in modified
symmetric teleparallel gravity, light and dark blue regions are constraints
in combination with CMB lensing reconstruction and BAO from Planck data. We
choose three different values of $p$, the black line for $p=3$, the red line
represents $p=4$ and the blue line for $p=5$.}
\label{fig:2}
\end{figure}

In Fig. \ref{fig:2}, we consider the hilltop inflation, where the
inflationary parameters are linked to $f(Q,T)$ gravity. It is apparent from
the calculations that the constant roll parameter has negligible effects on
the $\left( r,n_{s}\right) $ behavior, the plot shows that increasing the
value of $p$ makes the decreasing function obtained from our model
consistent with the latest observations provided by Planck's results. 
\begin{table}[h]
\centering%
\begin{tabular}{c|c|c|c}
\hline
\textrm{$\beta $} & \textrm{$p$} & \textrm{$N$} & \textrm{$r$} \\%
[0.5ex] \hline\hline
$-1.5$ & $6$ & $50$ & $0.32$ \\ \hline
$-1.1$ & $5$ & $55$ & $0.19$ \\ \hline
$-0.5$ & $4$ & $60$ & $0.10$ \\ \hline
$0.7$ & $3$ & $65$ & $0.049$ \\ \hline\hline
\end{tabular}%
\caption{Testing the tensor-to-scalar ratio for different parameters for the
hilltop inflation.}
\label{table:1}
\end{table}

Based on the previously established equation Eq.(\ref{r(N)}), we provide an
examination of the tensor-to-scalar ratio taking into account different
inflation durations that varies from $50$ to $65$ e-folds. In addition to an
increasing values of the constant roll $\beta $ parameter, the
tensor-to-scalar ratio $r$ can be compatible with the observational bound
for specific values of $p$\ and $N$. Furthermore, the constant roll
parameter and the inflationary e-folds provide compatible results only for
higher values since they are inversely proportional to $r$.

Now as we construct a constant roll model for the $f(Q,T)$\ gravity in the
context of chaotic and hilltop inflation, we will study the evolution of
primordial black holes that were supposed to occur just after the inflation
period taking into account our $f(Q,T)$\ gravity model.

\section{Primordial black holes evolution in $f(Q,T)$ Gravity}

\label{secIV}

\subsection{Rotating and non-rotating Primordial Black Holes}

Primordial black holes abundance is determined by the primordial power
spectrum. In fact, PBHs were formed as a result of amplification in the
primordial power spectrum on small scales at the time when primordial
inhomogeneities re-enter the Hubble horizon in a radiation-dominated
Universe era. Some regions with a significant positive curvature which are
considered equivalent to a closed Universe will collapse into a black hole 
\cite{D1}. At first, black holes were considered to be eternal and evolved
with an increasing mass by absorbing more matter or even other stars and
BHs. However, studying their quantum properties shows the possibility of
emitting particles with a thermal spectrum related to BHs surface gravity 
\cite{D2,D3}. In this process of emitting particles, BHs lose mass and
angular momentum with different\ properties depending on the specific
characteristics of BHs. In this direction, we will discuss the thermal
properties of evaporating PBHs for two cases namely the Schwarzschild and
Kerr PBHs.

As an analogy to non-rotating Schwarzschild black holes, we consider a PBH
with mass $M_{BH},$ the thermal spectrum of emitted particles from the
evaporation process has the following expression \cite{D4}%
\begin{equation}
T_{BH}=\frac{1}{8\pi GM_{BH}}\sim 1.06\left( \frac{10^{10}kg}{M_{BH}}\right)
GeV,  \label{Eqi1}
\end{equation}

Another option is that the evaporating BHs have some angular momentum, such
spinning BHs, also known as Kerr BHs, might have originated with an initial
spin or acquired their angular momenta by a variety of events, such as
mergers \cite{D5,D6,D7}. A PBH temperature for the Kerr scenario can be
modified due to their spinning and it's given as \cite{D8} 
\begin{equation}
T_{BH}=\frac{1}{4\pi GM_{BH}}\frac{\sqrt{1-a_{\ast }^{2}}}{1+\sqrt{1-a_{\ast
}^{2}}},  \label{Eqi2}
\end{equation}%
where $a_{\ast }$ is a dimensionless parameter bounded by the interval $%
0\leq a_{\ast }\leq 1.$\ The primary distinction between Kerr $\left(
a_{\ast }\neq 0\right) $ and Schwarzschild $\left( a_{\ast }=0\right) $ BHs
is that Kerr BHs are axially symmetric rather than spherically symmetric.
When $a_{\ast }$ grows, the emission of particles with angular momentum
spinning in the same direction as the BH is enhanced \cite{D9}. For the case 
$a_{\ast }=1$\ a BH would have $T_{BH}=0,$ this is traditionally prohibited
in any statistical system.\ Furthermore, its horizon would have vanished
revealing a bare space-time singularity and contradicting the Cosmic
Censorship Conjecture \cite{D9,D10,D11}.

\subsection{Primordial Black Holes evolution}

Since we are interested in studying the rate of mass loss from PBH, we
recall the process which reduces the mass of the black hole due to Hawking
evaporation which is defined by \cite{D12,D13}%
\begin{equation}
\left( \frac{dM}{dt}\right) _{eva}=-\frac{\hbar c^{4}}{G^{2}}\frac{\alpha
^{\prime }}{M_{BH}^{2}}.  \label{Eqi3}
\end{equation}%
where $G$\ is Newton's gravitational constant, $\hbar $ is the Planck
constant, $c$ is light speed, $\alpha ^{\prime }$\ is the spin parameter of
the emitting particles, and the\ black hole radius is given by $r_{BH}=2GM.$
After integration Eq. (\ref{Eqi3}) we obtain the evolution of PBH mass as%
\begin{equation}
M_{eva}=M_{i}\left( 1-\frac{t}{t_{eva}}\right) ^{\frac{1}{3}},  \label{Eqi4}
\end{equation}%
we can define the Hawking evaporation time scale $t_{eva}$ as%
\begin{equation}
t_{eva}=\frac{G^{2}}{\hbar c^{4}}\frac{M_{i}^{3}}{3\alpha ^{\prime }},
\label{Eqi5}
\end{equation}%
according to Ref. \cite{D14} fine-tuning the initial PBH masse $M_{i}$\
along with the $\alpha ^{\prime }$\ parameter would be a good method to
probe the early Universe. On the other hand, additional contributions
suggest that the evaporation of PBHs would have interesting implications on
the CMB and the standard cosmological nucleosynthesis scenario \cite{D15,D16}%
. 
\begin{figure}[H]
\centering
\includegraphics[width=15cm]{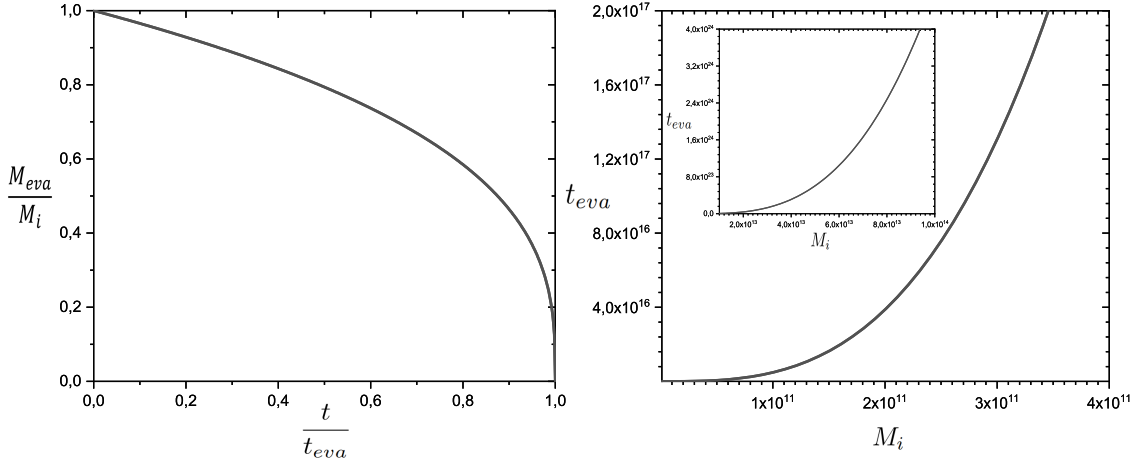}
\caption{On the left, the evolution of the primordial black hole reduced the
mass ratio to the initial mass $M_{eva}/M_{i}$ as functions of time ratio $%
t/t_{eva}$ is plotted.\ On the right, the evaporation time of primordial
black holes\ as a function of the initial mass is presented, and the plot in
the left corner shows the variation of $t_{eva}$\ for higher values of $%
M_{i}.$ \ }
\label{fig:3}
\end{figure}
In Fig. \ref{fig:3} we test the evolution of the primordial black holes
evaporation and initial mass ratio $M_{eva}/M_{i}$ with respect to the ratio 
$t/t_{eva},$ and we\ plotted the variation of the evaporation time parameter 
$t_{eva}$ as functions of the initial mass of primordial black holes\ $%
M_{i}. $\ Our results show that the evaporation mass can simply decrease as
we move forward in time. However, one must study the effect of the initial
primordial black holes masses on the time that must take in order to
completely evaporate. In this direction, the second plot indicates that as
we increase the initial mass of PBHs we need more time to achieve a complete
evaporation process. In fact, an initial mass in the order of $M_{i}\gtrsim
10^{13}kg$\ needs time more than the current age of our Universe to
evaporate, as a result, we must consider lower bounds on PBHs initial mass
for fine consistency with cosmic history.

The accretion of cosmic fluid surrounding the black hole will prolong the
evaporation of the primordial black holes. Therefore, it is necessary to add
a mass accretion rate which for a cosmic fluid with $\rho _{eff}$ and $%
p_{eff}$ will be given as%
\begin{equation}
\left( \frac{dM}{dt}\right) _{accr}=\frac{16\pi G^{2}}{c^{5}}M^{2}\left(
\rho _{eff}+p_{eff}\right) ,  \label{Eqi6}
\end{equation}%
which can be integrated to be in the final form \cite{D17}%
\begin{equation}
M_{accr}=M_{i}\left( 1-\frac{t}{t_{accr}}\right) ^{-1},  \label{Eqi7}
\end{equation}%
where from Eqs. (\ref{EqiX}) and (\ref{EqiY}) the accretion time scale $%
t_{accr}$ is defined as%
\begin{equation}
t_{accr}=\left[ \frac{16\pi G^{2}}{c^{5}}M_{i}\left( \rho
_{eff}+p_{eff}\right) \right] ^{-1}.  \label{Eqi8}
\end{equation}%
Additionally, we can write the following equation

\begin{equation}
\rho _{eff}+p_{eff}=\frac{3}{\kappa }H^{2}\left( \frac{2\left( \sigma
+\kappa \right) \left( 1+\omega \right) }{2\kappa -\left( \omega -3\right)
\sigma }\right)  \label{Eqi9}
\end{equation}

Here, at a radiation-dominated era, and using the fact that the Hubble
parameter is given by $H=\frac{\dot{a}}{a},$ we can simply choose $%
a(t)\propto t^{1/2}$ and finally obtain $H=1/\left( 2t\right) ^{2}$ to
estimate the time at which the radiation surrounded primordial black-holes. 
\begin{figure}[H]
\centering
\includegraphics[width=15cm]{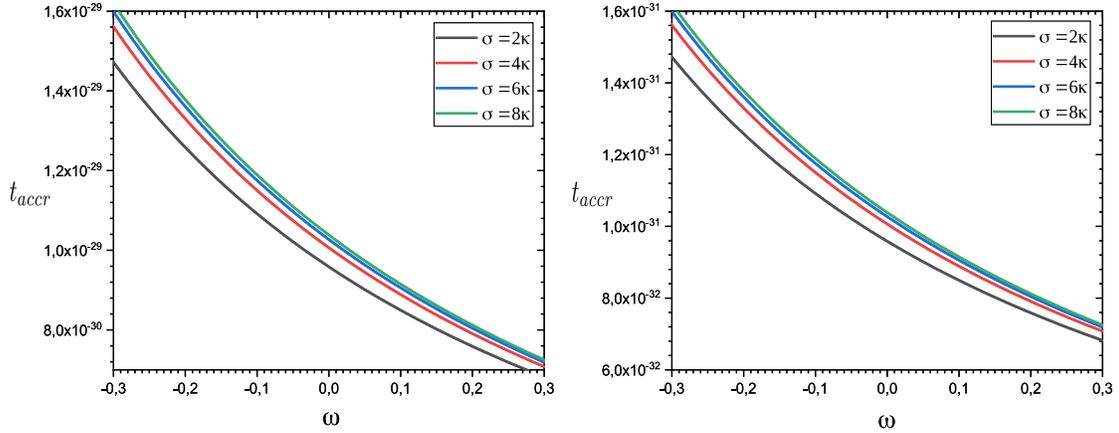}
\caption{The accretion time $t_{accr}$ variation as a function of the
equation of state in the interval $\protect\omega \in \left[ -0.3,0.3\right] 
$ considering different values of the $f(Q,T)$ gravity model parameter $%
\protect\sigma $. The left panel provides the case for $M_{i}\propto
10^{12}kg,$\ while the right shows the case of $M_{i}\propto 10^{14}kg$. }
\label{fig:4}
\end{figure}
Fig. \ref{fig:4} illustrates the evolution of the accretion time as a
function of a specific interval of the equation of state parameter. In our
study, the accretion time is the moment at which primordial black holes
started the accretion process, the variation of the accretion time shows to
be minimally\ dependent on the choice of $\sigma $\ parameter for our chosen 
$f(Q,T)$ gravity model. In fact, the behavior of accretion time increases
for higher values of $\sigma $.\ Considering the fact that in order for
inflation to initiate the EoS must be bounded by $\omega >-1/3.$ Moreover, $%
\omega $ evolves toward $1/3$ for inflation to be ended and the subsequent
periods to take place \cite{D18,D19}. In Eq.(\ref{Eqi9}) we choose a fixed
value of time using the Hubble parameter which represents the time of PBHs
formation, following the results of \cite{D12} which suggests that PBHs
formed at $t\propto 10^{-23}s$, our results provide precise values of the
time of accretion which decreases as we consider higher values of the EoS
parameter. On the other hand, for lower initial PBHs masses, the accretion
initiates sooner than in the case of higher $M_{i}$\ values. 
\begin{figure}[H]
\centering
\includegraphics[width=15cm]{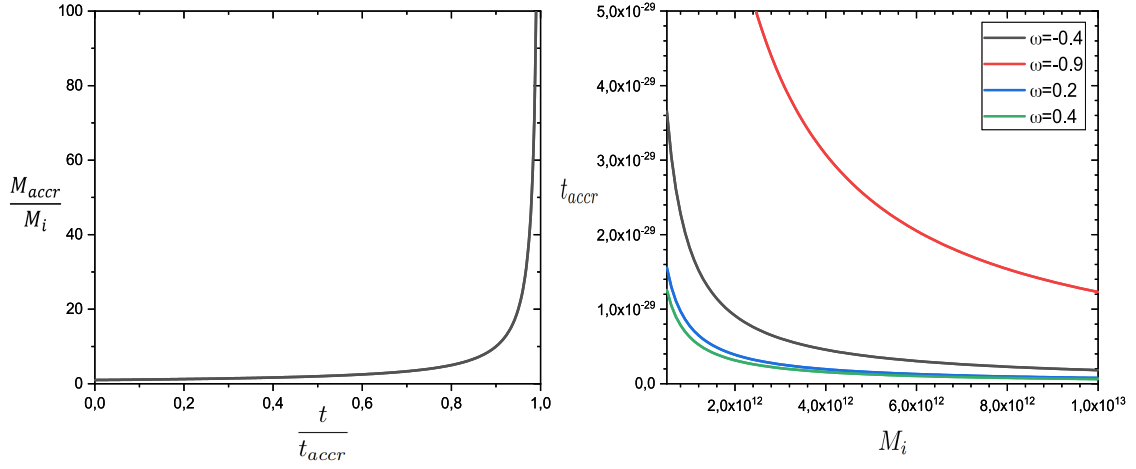}
\caption{On the left panel, the evolution of the primordial black holes
accretion mass ratio to the initial mass $M_{accr}/M_{i}$ as functions of
the time ratio $t/t_{accr}$ is shown.\ On the right one, the accretion time
of primordial black holes\ as a function of the initial mass is presented
for specific values on the equation of state $\protect\omega $\ parameter.}
\label{fig:5}
\end{figure}
In Fig. \ref{fig:5} we provide the variation of the primordial black holes
accretion mass ratio to the initial mass $M_{accr}/M_{i}$ as a function of
the ratio $t/t_{accr}.$ From this figure we conclude that when $%
t\longrightarrow t_{accr}$\ the resulting primordial black hole mass due to
the accretion process became in the order of $100M_{i}.$\ On the other hand,
accretion occurs faster when we consider more significant initial PBH masses
for several values of $\omega $ parameter.

The total PBHs mass evolution may now be rewritten in the following form $%
M_{BH}=M_{eva}+M_{accr}$%
\begin{table}[h]
\centering%
\begin{tabular}{c|c|c|c|c|c}
\hline
\textrm{$\sigma $} & \textrm{$\omega $} & \textrm{$t(s)$} & \textrm{$%
M_{i}(kg)$} & \textrm{$M_{BH}(kg)$} & \textrm{$T_{BH}(GeV)$} \\%
[0.5ex] \hline\hline
$1.5~\kappa $ & $-0.3$ & $10^{-15}$ & $10^{6}$ & $\sim 2\times 10^{6}$ & $%
\sim 5.3\times 10^{3}$ \\ \hline
$2~\kappa $ & $-0.2$ & $10^{-5}$ & $10^{7}$ & $\sim 1.9\times 10^{7}$ & $%
\sim 5.3\times 10^{2}$ \\ \hline
$2.5~\kappa $ & $-0.1$ & $10$ & $10^{8}$ & $\sim 1.9\times 10^{8}$ & $\sim
53 $ \\ \hline
$3~\kappa $ & $0$ & $10^{5}$ & $10^{9}$ & $\sim 1.9\times 10^{9}$ & $\sim
5.3 $ \\ \hline
$3.5~\kappa $ & $0.1$ & $10^{9}$ & $10^{10}$ & $\sim 1.9\times 10^{10}$ & $%
\sim 53\times 10^{-2}$ \\ \hline
$4~\kappa $ & $0.2$ & $10^{13}$ & $10^{11}$ & $\sim 1.9\times 10^{11}$ & $%
\sim 53\times 10^{-3}$ \\ \hline
$4.5~\kappa $ & $0.3$ & $10^{17}$ & $10^{12}$ & $\sim 1.9\times 10^{12}$ & $%
\sim 53\times 10^{-4}$ \\ \hline\hline
\end{tabular}%
\caption{Testing the evolution of the total non-rotating PBHs masses and
temperature in $f(Q,T)$ gravity.}
\label{table:2}
\end{table}
\begin{table}[h]
\centering%
\begin{tabular}{c|c|c|c|c|c|c}
\hline
\textrm{$\sigma $} & \textrm{$\omega $} & \textrm{$t(s)$} & \textrm{$%
M_{i}(kg)$} & \textrm{$a_{\ast }$} & \textrm{$M_{BH}(kg)$} & \textrm{$%
T_{BH}(GeV)$} \\[0.5ex] \hline\hline
$1.5~\kappa $ & $-0.3$ & $10^{-15}$ & $10^{6}$ & $0.1$ & $\sim 2\times
10^{6} $ & $\sim 5.28\times 10^{3}$ \\ \hline
$2~\kappa $ & $-0.2$ & $10^{-5}$ & $10^{7}$ & $0.3$ & $\sim 1.9\times 10^{7}$
& $\sim 5.17\times 10^{2}$ \\ \hline
$2.5~\kappa $ & $-0.1$ & $10$ & $10^{8}$ & $0.5$ & $\sim 1.9\times 10^{8}$ & 
$\sim 49.1$ \\ \hline
$3~\kappa $ & $0$ & $10^{5}$ & $10^{9}$ & $0.7$ & $\sim 1.9\times 10^{9}$ & $%
\sim 4.41$ \\ \hline
$3.5~\kappa $ & $0.1$ & $10^{9}$ & $10^{10}$ & $0.9$ & $\sim 1.9\times
10^{10}$ & $\sim 32.1\times 10^{-2}$ \\ \hline
$4~\kappa $ & $0.2$ & $10^{13}$ & $10^{11}$ & $0.96$ & $\sim 1.9\times
10^{11}$ & $\sim 23.1\times 10^{-3}$ \\ \hline
$4.5~\kappa $ & $0.3$ & $10^{17}$ & $10^{12}$ & $0.99$ & $\sim 1.9\times
10^{12}$ & $\sim 13.1\times 10^{-4}$ \\ \hline\hline
\end{tabular}%
\caption{Testing the evolution of the total rotating PBHs masses and
temperature in $f(Q,T)$ gravity.}
\label{table:3}
\end{table}

From tables (\ref{table:2}) and (\ref{table:3}) we study the evolution of
rotating and non-rotating PBHs masses and temperature as functions of
different parameters. According to these results, we conclude that PBHs
masses increase with higher initial masses $M_{i}$. On the other hand, PBHs
temperature decreases as we move forward in time. Moreover, $T_{BH}$\ is
inversely proportional to PBHs masses which makes the temperature evolve
from $TeV$\ to\ $MeV$\ taking into account masses in the order of $10^{12}kg.
$\ Finally, when we compare rotating and non-rotating black holes, we find
that higher values of the parameter $a_{\ast }$\ can simply lead to a lower
temperature for the case of rotating PBHs.

\section{Concluding Remarks}

\label{secV}

Over the past few decades, numerous studies have been conducted to examine
the early and late evolution of the Universe. The standard model of
cosmology, based on general relativity (GR), has proven to be a reliable
model for describing the dynamics of the Universe. However, there are still
some unresolved issues, such as the flatness and horizon problems, that need
to be addressed through further research on cosmological inflation.
Additionally, while GR has been successful in predicting the behavior of the
Universe, it is unable to fully explain the influence of dark sectors on its
dynamics in a way that is consistent with observed data. As a result, it may
be worthwhile to consider alternative models of gravity such $f(Q,T)$ model
to address these challenges.

In this study, we examined constant-roll inflation in the context of $f(Q,T)$
gravity. To do this, we started by explaining the basic theory of
cosmological inflation using the isotropic and homogeneous inflaton scalar
field. We then assumed a flat FLRW spacetime and an equation of state $%
\omega ,$ we provided a new technique for studying the constant-roll process
and correlating the slow-roll equations to the constant-roll $\beta $\
parameter based on the modified Friedmann equations obtained through $f(Q,T)$
gravity. Furthermore, we have calculated the inflationary observables, the
spectral index $n_{s},$ and the tensor-to-scalar ratio $r$ for two cases of
inflationary potentials, namely the chaotic and hiltop models. We showed
that for\ each model, a bound on the constant-roll parameter is preferred.
In the case of chaotic inflation, for a consistent value of $r$ and $n_{s},$ 
$\beta $\ must be bounded as $\beta \leq 0.$\ While for the Hiltop
Inflation, several parameters are involved to reproduce compatible values of 
$r$ and $n_{s},$ for instance, we must consider the bounds $p\geq 3$, $N\geq
60,$\ and $\beta \geq -0.5$ for a good index spectral and tensor-to-scalar
ratio\ consistencies.

Finally, for the PBHs evolution in the context of $f(Q,T)$ gravity, we
analyzed the accretion process and the evaporation through Hawking
radiation. From the obtained results, we conclude that both the evaporated
mass and the evaporation time are directly related to the initial mass $M_{i}
$ that must be bounded as $M_{i}<10^{13}kg$\ to be able to completely
evaporate currently or much earlier in the cosmic time. For the accretion
process, we can summarise that the accretion of matter and radiation are
model dependent in the context of the $f(Q,T)$ gravity, which motivates us
to explore the PBHs evolution in the framework of modified gravity.
According to the results, if we supposed that PBHs formed at $t\propto
10^{-23}s$ the PBHs mass due to the accretion can reach $100M_{i}$\ taking
into account that for lower initial PBHs masses the accretion can occur
faster. Lastly, we studied the PBHs evaporation taking into account the
effect of several parameters, and concluded that the Hawking temperature can
simply decrease for higher initial masses through the cosmic time for both
rotating and non-rotating black holes.

\end{document}